\documentclass[12pt,endfloat]{article}
\usepackage{varioref}
\usepackage{graphicx}
\usepackage[T1]{fontenc}
\usepackage{epsfig}
\usepackage{subfigure}
\usepackage{cite}
\usepackage[labelsep=space]{caption}
\usepackage{amsmath}
\usepackage{amsfonts}
\usepackage{stmaryrd}
\usepackage{psfrag}
\usepackage[a4paper]{geometry}
\usepackage{setspace}
\usepackage{color}
\usepackage{tabularx}\newcolumntype{Y}{>{\centering\arraybackslash}X}

\begin{document}


\title{Revealing photonic Symmetry-Protected Modes ($SPM$) by Finite-Difference-Time-Domain method}
\author{A. Hoblos$^{1}$, M. Suarez$^{1}$, B. Guichardaz$^1$, N. Courjal$^{1}$, \\M.-P. Bernal$^{1}$ and F. I. Baida$^{1*}$\\
$^1$ Institut FEMTO-ST, UMR CNRS 6174, Universit\'e Bourgogne Franche-Comt\'e\\ 25030 Besan\c con, France\\ 
$^*${Corresponding author: fbaida@univ-fcomte.fr}}
\maketitle
\begin{abstract}
This letter is devoted to point out a specific character of the Finite-Difference-Time-Domain method through the study of nano-structures supporting geometrical symmetry-protected modes that can not be excited at certain conditions of illumination. The spatial discretization performed in the FDTD algorithm naturally leads to break this symmetry and allows the excitation of these modes. The quality factor of the corresponding resonances are then directly linked to the degree of the symmetry breaking i.e. the spatial grid dimension even though the convergence criteria of the FDTD are fulfilled. This finding shows that the FDTD must be handled with a great care and, more importantly, that very huge quality-factor resonances could be achieved at the cost of nanometer-scale mastered fabrication processes. 
\end{abstract}

\section{Introduction}

The FDTD is one of the most used methods to model the interaction between light and matter at the nanoscale \cite{Taflove2005,sukharev:opn11}. Its use requires a minimum of knowledge in the field of electromagnetism. Several mistakes or misinterpretations have been made mainly when commercial codes are used. Worse than that, it happens that the developers themselves do the same. For example, let us cite our own study \cite{baida:apb07} in which we misinterpreted some results as it will be detailed in the following. Here, we will present recent results demonstrating an interesting character of the FDTD. The latter involves the direct definition of the structure under study in the real space without the need of Fourier Transform the object profile or geometry. Due to symmetry considerations, some eigenmodes of a given structure can not be excited in specific conditions of illumination. These modes are qualified by Symmetry Protected Modes or Bound states In the Continuum (BIC)\cite{vonNeumann29}. They could emerge as a consequence of destructive interference between two or more modes whose properties depend on the angle of incidence \cite{azzam:prl18}. Let us cite the Transverse ElectroMagnetic ($TEM$) guided mode inside metallic annular apertures that requires an oblique incidence to be excited \cite{baida:apb07}. In this last paper, the amplitude of the $TEM$ mode in the spectra of figure 3 seems to grow with the angle of incidence (see the peaks indicating by vertical black arrows). This results is a blatant artifact due to the decreasing of the quality factor of the resonance when the angle of incidence increases. In other words, all those transmission peaks must reach $100\%$ but need more time FDTD simulation. Recently, because of improvement in nanotechnology, this kind of modes has found a great expansion in the design of structure with very high quality factor resonances. In fact, as small as the breaking of the symmetry is, as large as the Q-factor of the resonance of the mode will be. Thus, two obvious solutions exist to break the symmetry: an intrinsic one by modifying the geometry of the structure itself \cite{ndao:apl13,alaridhee:oe15,foley:ol15,campione:acs06,Sadrieva:acsphot17} or an extrinsic one induced by modifying the illumination conditions (oblique incidence for instance)\cite{baida:apb07,ndao:jo14,yoon:sr15}. Two-dimensional (2D) \cite{foley:ol15} or three-dimensional (3D) \cite{he:sr15,lu:np16,campione:acs06,yuan:jpb17,kodigala:nature17} structures were proposed in this context to present symmetry protected modes at normal incidence. The spatial discretization exercised in the FDTD combined with the Yee scheme\cite{yee:ieee66}, for which two different components of the electromagnetic field are located at two different spatial positions, leads to naturally break the geometrical symmetry of the structure. By the way, these modes ($SPM$) are numerically revealed even in conditions where they are symmetry-protected. Nevertheless, finite element method (FEM), which is also based on a spatial discretization of the object, does not seem to prevent the existence of this type of symmetry-protected resonances \cite{cui:sr16} probably as a result of the mesh refinement generally used in FEM.

\section{Proposed structure and Results}

\begin{figure*}[!h]
\centering
\fbox{\includegraphics[width=0.9\textwidth]{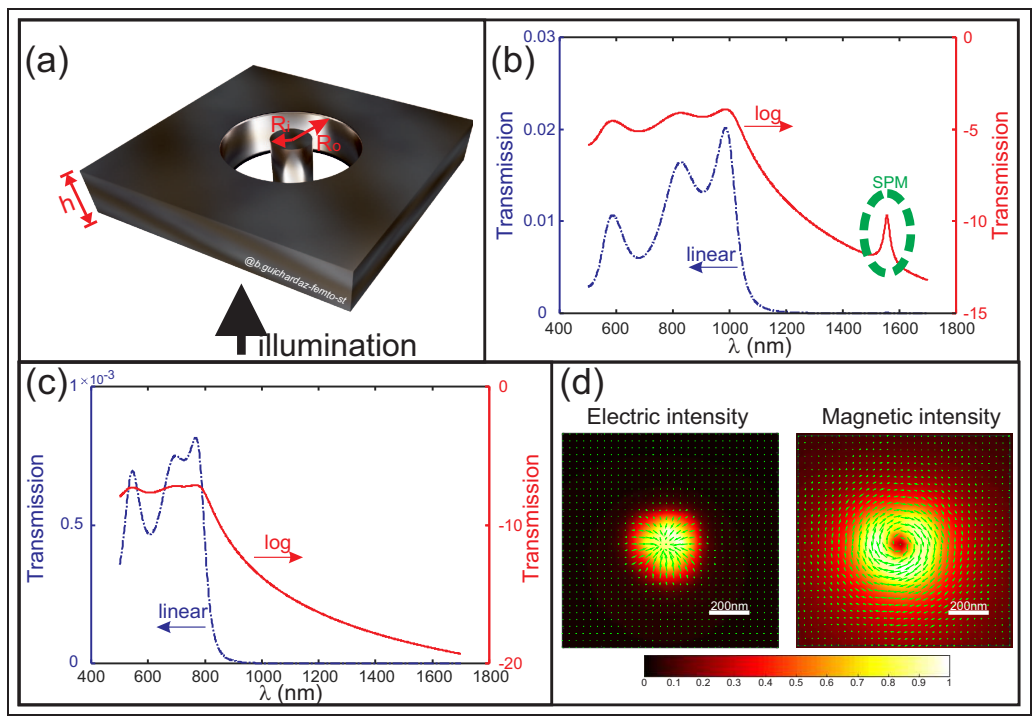}}
\caption{(a) Schema of the single coaxial aperture engraved into a self-suspended perfectly electric conductor layer. The inner and outer radii are fixed to $R_i=70~nm$ and $R_o=175~nm$ respectively and the metal thickness is $h=525~nm$. (b) Transmission spectra in linear scale (blue dashed-dotted line) and Logarithmic scale (red solid line) in the case of (b) 3D-Cartesian mesh and (c) 3D-cylindrical mesh. (d) Electric and magnetic field intensity distributions $70~nm$ up from the output side of the aperture calculated at $\lambda=1555.5~nm$ corresponding to excitation of the $SPM$ of figure (b) that occurs in Cartesian coordinates. The green arrows correspond to the transverse field component (in the $xOy$ plane).}
\label{coax_3DvsBOR}
\end{figure*}

Basically, the most adapted structures to have symmetry protected modes are structures with high degree of symmetry along 1D (slits or grooves) \cite{foley:ol15} or along 2D \cite{kodigala:nature17} with mirror symmetry or center of symmetry. Nonetheless, two completely different cases should be distinguished: the case of a localized resonance (LR) and the case of a collective resonance (CR). The first one can exist with only one pattern while the second one, comparable to a Fano-type resonance, requires a periodic structure with large number of patterns. To illustrate the case of a LR, we consider the coaxial aperture presented on figure \ref{coax_3DvsBOR}a. The fundamental guided mode inside the coaxial waveguide made in perfect electric conductor (PEC) is the $TEM$ (Transverse ElectroMagnetic) mode which has no cutoff-frequency. 

\begin{figure*}[!h]
\centering
\fbox{\includegraphics[width=0.9\textwidth]{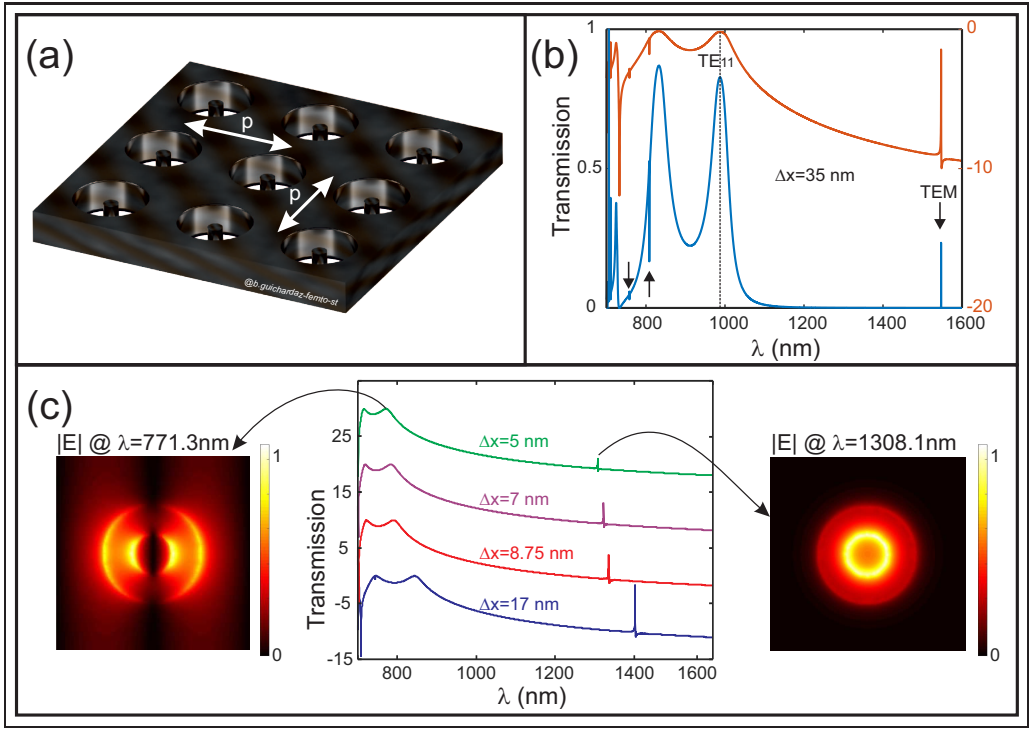}}
\caption{(a) Schema of the coaxial aperture grating engraved into a self-suspended perfectly electric conductor layer of $525~nm$ thickness. The period was fixed to $p=700~nm$ and the inner and outer radii are fixed to $R_i=70~nm$ and $R_o=175~nm$ respectively. (b) Transmission spectra in linear (blue solid line) and Logarithmic (dashed red line) scales calculated by FDTD with a mesh cell of $\Delta =35~nm$. (c) Transmission spectra in Logarithmic scale for four different FDTD mesh cell sizes varying from $\Delta=17.5~nm$ to $\Delta =5~nm$. The spectra are vertically shifted by a value of $10$ in order to enhance the visibility of all the peaks. The two color maps show the spatial distribution of the electric field amplitude $10~nm$ above the output side of the aperture of the $TE_{11}$ allowed guided mode (left) and of the $TEM/SPM$ mode (right) respectively. }
\label{3Dper_coax}
\end{figure*}

\begin{figure*}[!h]
\centering
\fbox{\includegraphics[width=0.9\textwidth]{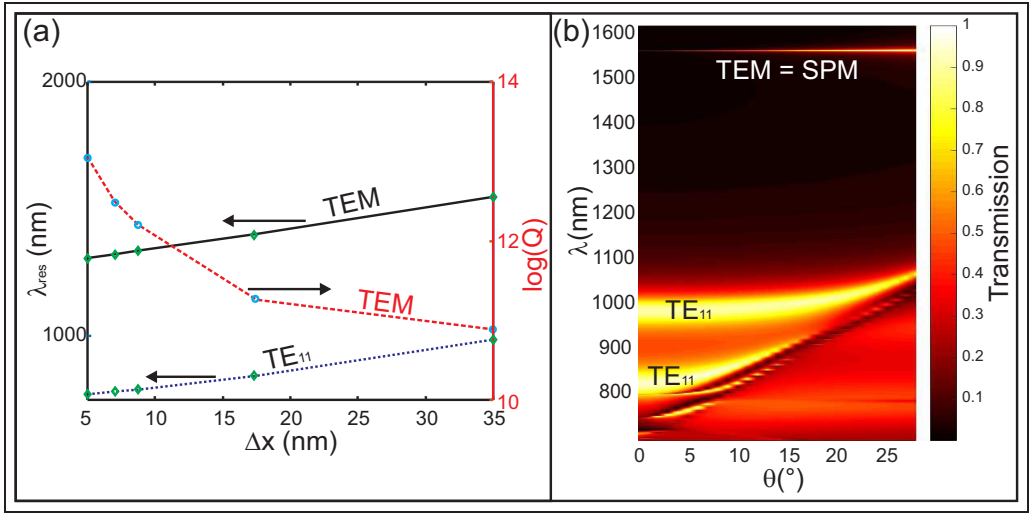}}
\caption{(a): Evolution of the spectral position of the symmetry-protected $TEM$ mode (black solid line) and the allowed guided TE$_{11}$ mode (blue dotted line) with respect to the FDTD cell size. The red dashed line corresponds to the modifications of the $SPM$ mode quality factor $Q$ plotted in Logarithmic scale. (b): Dispersion diagram of the annular aperture array calculated by SFM-FDTD method \cite{belkhir:pre08}.}
\label{evol}
\end{figure*}

When the spatial description of the structure retains its symmetry of revolution, this mode becomes a $SPM$ for a plane wave impinging the structure at normal incidence (along the waveguide axis). This can be numerically fulfilled if we describe the structure in cylindrical coordinates. Contrarily, if the latter is described within a cuboid mesh grid in Cartesian coordinates, the symmetry of the structure is therefore numerically broken and the $TEM$ mode could be excited. Figures \ref{coax_3DvsBOR}b,c depicted the two transmission spectra (normalized transmitted Poynting flux) calculated by 3D-Cartesian FDTD and by Body-Of-Revolution FDTD homemade codes respectively. The geometrical parameters of the aperture are given in the figure caption and the spatial grid was fixed to $\Delta=35nm$ in both simulations. As expected, the signature of the $SPM$ excitation only appears in the case of 3D-Cartesian mesh (Fig. \ref{coax_3DvsBOR}b). Its spectral position (here $\lambda_{SPM}\simeq1555~nm$) depends on the length of the waveguide (i.e. the metal film thickness) through a phase-matching condition \cite{baida:prb06}. Note that these two simulations were done over an identical total time of light-matter interaction, i.e. $53.4ps$ ($1.14\times 10^6$ time steps in Cartesian coordinates and $0.912\times 10^5$ time steps in cylindrical coordinates). To demonstrate the cylindrical symmetry of this $SPM$ mode, the electric intensity distribution is calculated in monochromatic regime at $\lambda=1555.5~nm$ in the case of a cuboid-mesh of $\Delta =35~nm$. Figure \ref{coax_3DvsBOR}d show this distribution at $70~nm$ for the output side of the coaxial aperture in addition to the magnetic field one. The radial character of the electric field (see green arrows) and the ortho-radial one of the magnetic field are the signature of the $SPM$-mode ($TEM$-mode) excitation.     
\begin{table*}[!h]
\centering
\caption{Evolution of the $SPM$ properties : spectral position ($\lambda_{TEM}$), Transmission coefficient (TC$_{TEM}$) and quality factor (Q$_{TEM}$) calculated by Harmonic Inversion algorithm (Harminv) and by Fourier Transform (FT) as a function of the FDTD cell size. The last column gives the spectral position of the TE$_{11}$ allowed guided mode.} 
\begin{tabular}{|c|c|c|c|c|c||c|}
  \hline
     FDTD cell size  & \multicolumn{2}{c|}{ $\lambda_{TEM}$ (nm)} & TC$_{TEM}$& \multicolumn{2}{c|}{$Q_{TEM}$} & $\lambda_{TE}$ (nm) \\
		\cline{2-7}
	 $\Delta$ (nm) & Harminv & FT & FT & Harminv & FT& FT\\
	\hline
  35 & 1546.97 & 1546.97 & 0.455 & 53683 & 4222 & 987.5 \\
  17.5 & 1402.90 & 1402.92 &  0.2428 & 77648 & 4677 & 843.5\\
	8.75 & 1336.21 & 1336.21 &  0.0138 & 199104 & 4996 & 790\\
 7 & 1323.02 & 1322.97 &  1.037$\times 10^{-3}$ & 266887 & 5091 & 783.2\\
 5 & 1308.13& 1308.09  &3.07$\times 10^{-4}$ & 467995 & 5307 & 771.3\\
  \hline
\end{tabular}\label{table}
\end{table*}
On the other hand, if we consider a bi-periodic ($p_x=p_y=700~nm$) structure such presented in figure \ref{3Dper_coax}a with the same coaxial aperture pattern, we can note that the transmission spectrum depicted on figure \ref{3Dper_coax}b exhibits several very thin resonance peaks (see vertical black arrows). The latter are due to the excitation of the vertical cavity harmonics of the same $SPM$ mode (the $TEM$ mode) and can only be obtained at the coast of a very large number of time-steps (significant light-matter interaction time). The peak amplitude is very enhanced compared to the one of the single aperture of Fig. \ref{coax_3DvsBOR} due to the sub-wavelength periodic character of the structure (all transmitted energy is propagating along the normal incidence direction). Fundamentally, the considered structure has a central symmetry so that when a resonance is excited, the transmission peak amplitude must reach $100\%$ \cite{alaridhee:oe15} as long as it is located beyond the Rayleigh anomaly (here over all the considered spectral range). This is clearly in contradiction with the obtained spectra of Fig. \ref{3Dper_coax}b. The reason is still linked to the same origin which is the spatial discretization responsible for the break of symmetry. To be convinced, we present on Fig. \ref{3Dper_coax}c the transmission spectrum of the same structure calculated by FDTD for different mesh cell sizes. As previously mentioned, the same light-matter interaction time is considered in all simulations (here $13~ps$). The inset presented on the right of this figure shows the distribution of the electric field amplitude at the $SPM$ wavelength $10~nm$ from the output side of the coaxial aperture in the case of a mesh cell size of $\Delta=5~nm$. The circular symmetry of the electric field is clearly visible confirming the excitation of the $TEM$ mode inside the aperture. The left distribution corresponds to the TE$_{11}$ mode showing a mirror symmetry that is consistent with the linear polarization of the incident plane wave.    
All the transmission properties for the different FDTD cell sizes are summarized in Table 1.
On one hand, Table 1 shows that when the cell size varies, both the position and the amplitude (transmission coefficient) of the peaks are significantly modified. The quality factor was estimated from two different methods: (a) the "Harminv" value is evaluated using a Harmonic Inversion Algorithm \cite{harminv} widely associated with FDTD simulations to extract information on high quality factor resonances \cite{wolf:book14} and, (b) by simply Fourier Transform (FT) the temporal recorded signal by the FDTD. The latter always underestimates the real value of the quality factor as demonstrated in Table 1. More importantly, one can note that the Q-factor increases drastically when the cell size decreases. Fundamentally, it tends to infinity when $\Delta \rightarrow\infty$ for any $SPM$ mode.
On the other hand, the FDTD convergence criteria are fulfilled for all the cell sizes considered in our simulations. In fact, it is well known (see ref. \cite{Taflove2005}) that the spatial convergence criterion of the FDTD ($\Delta<\lambda/16$) is purely numerically determined by allowing a phase velocity error of the wave during its propagation smaller than $0.1\%$ \cite{Taflove2005}. The accuracy of FDTD results mostly depends on the good depiction of the structure geometry. Thus, in Nano-Optics, a mesh size as small as $\lambda/100$ is often used ($\lambda/83$ in ref. \cite{roussey:josab07} and $\lambda/775$ in ref. \cite{barakat:oe10}) to simulate light-matter interaction involving structures with fine details. Thus, for the $SPM$ mode (here the $TEM$ mode of the coaxial waveguide), we can see that, the finer the mesh, the more the spectral position of the peaks shifts towards the blue spectral region and their amplitude decreases. However, for this mode, the spectral position is fundamentally governed by a phase matching condition involving the metal thickness (see Eq. 1 in ref. \cite{alaridhee:oe15}). The latter becomes more and more precise as the mesh is refined. The blue shift is then the signature of a metal thickness decreasing when the mesh is enhanced. 
The mesh refining also affects the aperture radius values. In fact, the spectral position of the right wide peaks, corresponding to the cutoff wavelength of the TE$_{11}$ guided mode inside the coaxial aperture \cite{baida:apb04} (see Fig. \ref{3Dper_coax}b), is analytically given by $\pi(R_o+R_i)=770~nm$ [$R_o$ and $R_i$ being the outer and the inner radii of the coaxial aperture respectively]. The spectra of figure \ref{3Dper_coax}c show a blue shift of this value from $987.5~nm$ to $771.3~nm$ when the cell size varies from $\Delta=35~nm$ to $\Delta=5~nm$ indicating, once again, an overestimation of the geometrical parameters as the mesh is coarse. Figure \ref{evol}a shows the evolution of the spectral position of the $SPM$ mode ($\lambda_{res}$ in black solid line) and its quality factor $Q$ (in dashed red line) with respect to the cell size in addition to the spectral position of the TE$_{11}$ (blue dotted line). One can note that the spectral position of the two modes exhibit a quite linear behavior with identical slope meaning a similar origin, namely the spatial discretization of the structure.  
As previously mentioned, the amplitude of the $SPM$ transmission peak resonance is directly related to the symmetry breaking induced by the mesh. The larger the latter (coarse mesh), the more the peak becomes visible (larger amplitude and smaller quality factor). Its amplitude decreases from $TC=0.45$ for $\Delta=35~nm$ to $TC=3\times10^{-4}$ for $\Delta=5~nm$ (see Table 1). Note that the latter value of the transmission is obtained by FDTD simulation that corresponds to a time-matter interaction of $13~ps$ and can increase if we consider larger time. On the other hand, the peak amplitude corresponding to the allowed guided mode (here the TE$_{11}$ mode at $\lambda=770~nm$) increases reaching almost $100\%$ thanks to the improvement of the central symmetry of the structure.
To go further in the demonstration of the $SP$ character of the resonance, we propose to intrinsically break the symmetry of the structure by illuminating it under oblique incidence. The transmission diagram of figure \ref{evol}b is calculated through a SFM-FDTD algorithm \cite{belkhir:pre08} within the rough meshing ($\Delta=35~nm$). It clearly shows that the spectral position of the $SPM$ is quite independent from the angle of incidence while its quality factor tends to infinity at normal incidence. Consequently, the resonance obtained at normal incidence has a physical meaning even if it should not exist for a perfectly symmetrical structure. These numerical simulations made by FDTD obviously demonstrate its ability to take into account the presence of fabrication imperfections. In fact, currently, the manufacturing processes used in nano-fabrication exhibit an accuracy of around $a=5~nm$ \cite{fib:2019} which is insufficient to ensure absolute symmetry of the fabricated structure. These residual imperfections (roughness, wall conicity,...) can be exploited to generate very thin resonances. For the FDTD, the spatial discretization value ($\Delta$) could give in consideration the real fabrication imperfections (if $\Delta=a$) and lead to a flagrant disagreement with analytical theory even if it fulfills the spatio-temporal convergence criteria of the FDTD. In such cases, FDTD reveals its superiority and its ability to integrate, by default, the uncertainties on the geometry of the structures under study.

\section{Summary}

FDTD is a powerful tool for the modeling in Nano-Optics. Nonetheless, simulation results should be carefully interpreted. Respecting the FDTD convergence criteria is not a guarantee of obtaining quantitative physical results, but allows, through a coarse mesh, revealing certain hidden physical phenomena taking intrinsically into account the manufacturing defects.

\section*{Acknowledgements} Computations have been performed on the supercomputer facilities of the "M\'esocentre de calcul de Franche-Comt\'e". This work has been partially supported by the EIPHI Graduate School (contract ANR-17-EURE-0002). 

\end{document}